\documentclass[a4paper]{jpconf}
\usepackage{graphicx}
\begin{document}
\title{Coulomb correction to the spectral bremsstrahlung rate in
the quantum Migdal theory of the Landau-- Pomeranchuk effect}

\author{O Voskresenskaya}

\address{Joint Institute for Nuclear Research, 141980,
Dubna, Moscow region, Russia}

\ead{voskr@jinr.ru}

\begin{abstract}
High-energy Coulomb corrections (CCs) to some quantities of the quantum Migdal theory of the
Landau--Pomeranchuk--Migdal (LPM) effect
are obtained analytically and numerically for the regimes of small and strong LPM suppression on the basis of a refined  Moli\`{e}re multiple scattering theory. Numerical calculations are presented in the ranges of the nuclear charge number of a scatterer $6\leq Z \leq 92$ and the expansion parameter of Moli\`{e}re $4.5\leq B^B \leq 8.5$. It is shown that  the
relative CC to the spectral bremsstrahlung rate can reach a value of the order of $-19\%$ in the ranges considered and must be borne in mind, e.g., in the Monte-Carlo analysis of electromagnetic cascade  LPM showers in extremely high-energy region.
\end{abstract}

\section{Introduction}

\medskip

The theory of the multiple scattering of charged particles has been treated by several authors. However, the most widely used at present is the multiple scattering theory of Moli\`{e}re \cite{M}. The multiple scattering theory is of interest for numerous applications related to particle transport in matter \cite{N} and is widely used in the cascade shower theory \cite{KN}.
The neutrino-oscillation experiments \cite{Am}, the astronomical cosmic-ray observations \cite{Ao}, etc., face a problem of taking into account the multiple scattering effects.

As the Moli\`{e}re theory can be currently used in ultrahigh-energy region, the role of the high-energy CCs to the parameters of this theory becomes significant. Of especial importance is the Coulomb correction to the screening angular parameter, as this parameter  enters into other important quantities of the Moli\`{e}re theory and the Migdal theory of the LPM effect \cite{KVT, VKT}. In \cite{VKT} it is shown that the CC to the spectral bremsstrahlung rate of the classical LPM effect theory \cite{MC} for the regime of small LPM suppression allows completely eliminating the discrepancy between the predictions of the LPM effect theory and its measurement at least for high-$Z$ targets and also to further improve the agreement between the predictions of the LPM effect theory analogue for a thin layer of matter and data of SLAC E-146 experiment \cite{SLAC}.
The aim of the presented work is a preliminary estimation of CCs to some important quantities 
in the quantum LPM effect theory, 
especially to the Migdal function $G(s)$ and the spectral bremsstrahlung rate which are of special interest in describing the shower production at energies $E$ exceeding $10^{13}$ eV \cite{MQ,Klein}.

Qualitative evidence of the LPM effect for bremsstrahlung in nuclear-emulsion study
of cosmic-ray-produced electromagnetic cascades over the range
  $10^{11}\leq E\leq 10^{13}$ eV was first reported in ref.~\cite{Varf1}.~The quantitative confirmation of this LPM predicted
bremsstrahlung suppression at Serpukhov experiment gives \cite{Varf2}.~A direct evidence for reduction of the bremsstrahlung cross section due to the LPM effect in the measurements of cosmic ray electrons at $E\sim 10^{12}$ eV was found in \cite{YKN}.
  In ref.~\cite{G} it is point out a significant reduction of the Migdal cross section
  due to LPM effect for neutrino-induced electromagnetic showers in ice  over the range
  $10^{16}\leq E\leq 10^{20}$ eV; the other mechanisms of its reduction are specified in \cite{Klein3}.

The significance of LPM effect in studying the average behavior of electromagnetic cascade showers at extremely high energies (the so-called ``LPM showers'' \cite{Misaki1}) was shown 
for energies $E>3.5\cdot 10^{13}$ eV in lead \cite{Misaki1,Stanev}, $E>2.2\cdot 10^{15}$ eV in standard rock \cite{Misaki2}, $E>10^{17}$ eV in air \cite{Streit,Dedenko}, etc. The big difference between the averaged LPM shower and usual (``Bethe--Heitler'' \cite{Misaki1})
showers over the range $10^{15}\leq E\leq 10^{21}$ eV in lead, water and standard rock is explained in \cite{Misaki1,Misaki2}. A series of reports \cite{Conf1} is devoted to applications of the LPM effect to extremely high-energy cosmic ray research in the cosmic rays physics.

The first attempt to apply the LPM effect to the consideration of the electromagnetic cascade shower at superhigh energies  was carried out in \cite{Poman}.
The first correct results obtained for the LPM cascade showers through Monte-Carlo simulation techniques with taking into account the LPM effect for one-dimensional case presents \cite{First1}. For both the one-dimensional and the three-dimensional cases, they were given in \cite{First2}. Calculations of LPM showers by a hybrid method, which combines a Monte-Carlo simulation with analytical calculations, were performed in \cite{Stanev, Dedenko2}.
In \cite{First3} a matrix method for calculation of LPM showers was proposed.
Characteristics of individual cascade  LPM showers have been studied in \cite{KATM,Conf2}. There have been shown the multi-peak structure of individual LPM showers \cite{KATM} and a strong diversity among them \cite{Conf2}.

Careful investigation of fluctuations in LPM showers by simulation technique, which
takes into account the LPM effect, requires a sufficiently high accuracy of its theoretical description. Since the Migdal theory does not include all the corrections that should in principle
be included (in particular, the CCs) \cite{Cill}, the refinement of the quantities of the conventional Migdal LPM effect theory by means of the high-energy CCs is of considerable interest.
The present work is the first step in this direction. In Section 2
we present briefly our results for the high-energy CCs to the parameters of the Moli\`{e}re
multiple scattering theory \cite{KVT}. Then, in Section 3 we report the results of applying this
improved Moli\`{e}re multiple scattering theory for obtaining CCs to the quantities of the classical 
and quantum LPM effect theories by Migdal \cite{MC,MQ}.
Finally, in Sec. 4 we briefly sum up our results and discuss some perspectives.



\section{Coulomb corrections to the parameters of the Moli\`{e}re theory}

\medskip

In this section we  present an exact analytical result for the Moli\`{e}re  screening
angular parameter $\theta_a$ ($\theta_a^{\,\prime}=\sqrt{1.167}\theta_a$),
valid to all orders in the Born parameter $\xi=Z\alpha/\beta$,


\begin{equation}\label{basres}
\Delta_{\scriptscriptstyle
CC}[\ln\big(\theta_a^{\,\prime}\big)]\equiv
\ln(\theta^\prime_a)-\ln(\theta^\prime_a)^{\scriptscriptstyle B}
=f(\xi)\,
\end{equation}
instead of an approximate one for the Moli\`{e}re $\theta_a^{\scriptscriptstyle M}$ and
first-order Born $\theta_a^{\scriptscriptstyle B}$ values of screening angle, valid to second order in the parameter $\xi$,

\begin{equation}
\label{19}
\theta_a^{\scriptscriptstyle M}=\theta_a^{\scriptscriptstyle B}\sqrt{1+3.34\,
\xi^2}
\end{equation}
 that was obtained in the original paper of Moli\`{e}re \cite{M}. Here,
 $\alpha$ denotes the fine structure constant,  $\beta=v/c$ is the velocity of a projectile in units of the velocity of light, $\theta_a^{\scriptscriptstyle B}=1.20\,\alpha\, Z^{1/3}$ and
 $f(\xi)\equiv\xi^2\sum_{n=1}^\infty [n(n^2+\xi^2)]^{-1}$ represents the Bethe--Maximon function.

We also present the analytical and numerical results for the Coulomb
corrections to the parameters $b$, $B$ and $\overline{\vartheta^2}$ of the Moli\`{e}re
expansion method for the angular distribution function
\begin{equation}\label{power} w_{\scriptscriptstyle
M}(\vartheta,L)=\sum\limits_{n=0}^{\infty}\frac{1}{n!}\frac{1}{B^n}w_n(\vartheta,L),\quad
w_n(\vartheta,L) =
\frac{1}{\overline{\vartheta^{\,2}}}\int\limits_0^{\infty}y dy J_0
\left(\frac{\vartheta}{\overline{\vartheta}}\, y\right)
e^{-y^2/4}
\left[\frac{y^2}{4}\ln\left(\frac{y^2}{4}\right)\right]^n\ ,
\end{equation}
where $\vec\vartheta$ is a two-dimensional particle scattering
angle in the plane orthogonal to the incident particle direction, the relation $\overline{\vartheta^{\,2}}=\theta_c^2B$ defines the average square value of $\vec\vartheta$, $\theta_c$ is
the characteristic angle, $L$ denotes the thickness of an absorber, $y=\theta_c\eta$,
$\eta$ signifies  the Fourier--Bessel transform variable corresponding to $\vartheta$,  $J_0(\vartheta \eta)$ represents the
zero-order Bessel function of the first kind and the expansion parameter $B$ is defined by the transcendental equation $B-\ln B=b$ with $b=\ln(\theta_c/\theta_a^\prime)^2$.

For the Coulomb corrections to the Born parameters $b^{\scriptscriptstyle B}$, $B^{\scriptscriptstyle B}$ and ${\overline{\vartheta^2}}^{\scriptscriptstyle
B}$, we get
\begin{equation}\label{limcorrec1}
\Delta_{\scriptscriptstyle CC}[b]=-f(\xi),\; \Delta_{\scriptscriptstyle
CC}[B]=f(\xi)/(1/B^{\scriptscriptstyle B}-1),\;
\Delta_{\scriptscriptstyle CC}\!\left[\overline{\vartheta^2}\right]=\theta_c^2 \,\Delta_{\scriptscriptstyle CC}\!\left[B\right].
\end{equation}
The relative Coulomb corrections become
\begin{equation}\label{rel}
\delta_{\scriptscriptstyle CC}\!\left[\overline{\vartheta^2}\right]=\delta_{\scriptscriptstyle
CC}\!\left[B\right]
=f(\xi)/(1-B^{\scriptscriptstyle B}),\; \delta_{\scriptscriptstyle CC}[\theta_a]=\exp\left[f\left(\xi\right)\right]-1\ .
\end{equation}
The relative difference between the  approximate $\theta_a^{\scriptscriptstyle \mathrm{M}}$ and exact $\theta_a$ results reads
\begin{eqnarray}\label{angle}
\delta_{\scriptscriptstyle
\mathrm{CCM}}[\theta_a]
\equiv(\theta_a-
\theta_a^{\scriptscriptstyle \mathrm{M}})/
\theta_a^{\scriptscriptstyle \mathrm{M}}=\theta_a/
\theta_a^{\scriptscriptstyle \mathrm{M}}-1=R_{\scriptscriptstyle \mathrm{CCM}}[\theta_a]-1\,.
\end{eqnarray}

The $Z$ dependence of these corrections and differences presents figure~1. It shows that
while the modules of the quantities $\delta_{\scriptscriptstyle CC}\!\left[\overline{\vartheta^2}\right]$ and $\Delta_{\scriptscriptstyle
\mathrm{CCM}}[\theta_a]$
reach about $6\%$ for high-$Z$ targets
(dashdotted and dashed lines),
the maximal value of $\delta_{\scriptscriptstyle CC}[\theta_a]$  amounts
approximately to 50\% for $Z=92$ (upper full line). The modules of
CCs to the parameters $b$ and $B$ are: $\big\vert\Delta_{\scriptscriptstyle CC}[B]\big\vert\sim 0.45$ (lower full line) and $\big\vert\Delta_{\scriptscriptstyle CC}[b]\big\vert\sim 0.40$
(lower broken  line), such as $\Delta_{\scriptscriptstyle CC}[\ln\left(\theta_a^{\,\prime}\right)]\sim 0.40$ for $Z=92$ (upper broken  line).
These corrections are also valid for such modifications of the Moli\`{e}re theory as those proposed in \cite{N}. They will also be used below.

\begin{figure}[h!]

\begin{center}

\includegraphics[width=0.35\linewidth]{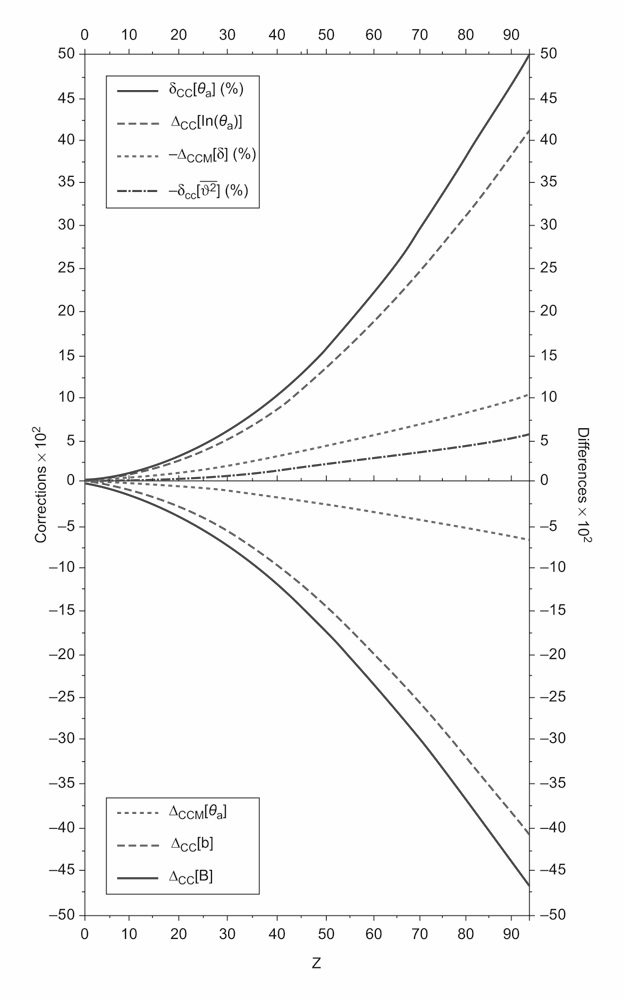}

\end{center}
\caption{\label{Figure1}The $Z$ dependence of the Coulomb corrections
to some parameters of the Moli\`{e}re multiple scattering theory and the differences
between exact and approximate results \cite{KVT}.}
\end{figure}

\newpage

\section{Applying the improved Moli\`{e}re theory to the description of the LPM effect}

\smallskip

\subsection{Coulomb corrections to the quantities of classical LPM
effect theory}

\smallskip

Based on the Coulomb corrections found in the previous section, we
first get the analytical and numerical results for the Coulomb corrections to the quantities of the  classical Migdal theory of the LPM effect for sufficiently thick targets, the basic formulae of which in the units  $\hbar = c =1$, $e^2=1/137$ read \cite{MC,ASh}
\begin{eqnarray}\label{7}
\left\langle\frac{dI}{d\omega}\right\rangle
&=&\Phi(s)\left(\frac{dI}{d\omega}\right)_0,\quad
\left(\frac{dI}{d\omega}\right)_0\:=\:\frac{2e^2}{3\pi} \gamma^2q\,L\ ,\\
\label{8} \Phi(s)&=&24s^2\left[\int\limits_{0}^{\infty}
dx\,e^{-2sx}\mbox{cth}(x)\sin(2sx)-\frac{\pi}{4} \right]\ ,\\
q&=&\overline{\vartheta^2}/L,\quad
s^2\:=\:\frac{\lambda^2}{\overline{\vartheta^2}},\quad \lambda^2\:=\:\gamma^{-2}\ .\nonumber
\end{eqnarray}
Here, $\left\langle dI/d\omega\right\rangle$ is  the electron spectral bremsstrahlung intensity averaged over various trajectories of electron motion, $\left(dI/d\omega\right)_0$
signifies the spectral bremsstrahlung intensity without accounting the multiple scattering effects  in the radiation and $\gamma$ denotes the Lorentz factor of the scattered particle.

The analytical result for the Coulomb correction $\Delta_{\scriptscriptstyle CC}$
to the Born spectral bremsstrahlung rate $(dI/d\omega)_0$ is as follows:
\begin{eqnarray}
\label{CCw0}
\Delta_{\scriptscriptstyle
CC}\!\left[\left(\frac{dI}{d\omega}\right)_0\right]
=\frac{2e^2}{3\pi} \gamma^2L\,\Delta_{\scriptscriptstyle CC}[q]
,\quad \Delta_{\scriptscriptstyle CC}[q]=\frac{1}{L}\,\Delta_{\scriptscriptstyle CC}\!\left[\overline{\vartheta^2}\right]\,,
\end{eqnarray}
where $\Delta_{\scriptscriptstyle CC}\!\left[\overline{\vartheta^2}\right]$ is given by (\ref{limcorrec1}).
In doing so, $\Delta_{\scriptscriptstyle
CC}\!\left[(dI/d\omega)_0\right]$ and
$\delta_{\scriptscriptstyle
CC}\left[(dI/d\omega)_0\right]$
become
\begin{equation}
\label{10}
\Delta_{\scriptscriptstyle
CC}\!\left[\left(\frac{dI}{d\omega}\right)_0\right]
=\frac{2(e\gamma\theta_c)^2}{3\pi\,(1/B^{\scriptscriptstyle
B}-1)}\, f(\xi),\quad
\delta_{\scriptscriptstyle
CC}\!\left[\left(\frac{dI}{d\omega}\right)_0\right]=
\frac{f(\xi)}{1-B^{\scriptscriptstyle B}}\ .
\end{equation}

For the regime of small  LPM suppression, $\Phi(s)\approx 1-0.012/s^4$, the analytical result for the relative Coulomb correction to the
Migdal function $\Phi(s)$ in the entire range $1\leq s \leq \infty$ reads

\begin{equation}
\delta_{\scriptscriptstyle \mathrm{CC}}\!\left[\Phi(s) \right]
=\frac{0.012}{s^4}\,\delta_{\scriptscriptstyle \mathrm{CC}}\!\left[s^4
\right] \frac{\left(s^4\right)^{\scriptscriptstyle \mathrm{B}}
}{\left(s^4\right)^{\scriptscriptstyle \mathrm{B}}-0.012}\ ,
\end{equation}
where
\begin{equation}
\label{12}
\delta_{\scriptscriptstyle \mathrm{CC}}\!\left[s^4
\right]=
1-\left(\frac{\big(\overline{\vartheta^2}\big)^{\scriptscriptstyle
\mathrm{B}}}{\overline{\vartheta^2}}\right)^2
=1-\frac{1}{{\left(\delta_{\scriptscriptstyle
\mathrm{CC}}\big[\overline{\vartheta^2}\big]+1\right)}^{2}}\,.
\end{equation}

Table~\ref{table1} presents the sample means ${\bar\delta}_{\scriptscriptstyle CC}\!\left[\left\langle dI/d\omega\right\rangle\right]$ $(\%)$ of the relative CCs for
some targets of the experiment \cite{SLAC} and also the population mean ${\delta}_{\scriptscriptstyle CC}\!\left[\left\langle dI/d\omega\right\rangle\right]_\mu\,(\%)$ over the entire range  $1.0 \leq s\leq \infty$ of the parameter $s$.
It  shows that $\bar\delta_{\scriptscriptstyle CC}\!\left[\langle
dI/d\omega\rangle\right]=(-4.50\pm 0.05)\%$ ($Z=82$) and
$\bar\delta_{\scriptscriptstyle CC}\![\left\langle
dI/d\omega\rangle\right]=(-5.35\pm 0.06)\%$ ($Z=92$) coincide within
the experimental error with the values of the normalization
corrections $(-4.5\pm 0.2)\%$ for $2\%L_{\scriptscriptstyle R}$ lead
target and $(-5.6\pm 0.3)\%$ for $3\%L_{\scriptscriptstyle R}$ uranium
target, respectively;
${\delta}_{\scriptscriptstyle CC}\!\left[\left\langle dI/d\omega\right\rangle\right]_\mu= (-4.70\pm 0.49)\%$ excellently agrees with the  weigh\-ted average  $(-4.7\pm 2)\%$ of
the normalization correction obtai\-ned for 25 GeV data of the SLAC E-146 experiment \cite{SLAC}.

This means that applying the improved multiple scattering theory by Moli\`{e}re allows one to
avoid multiplying the results of predictions of the conventional Migdal LPM effect theory
by a normalization factor and leads to agreement between
the improved Migdal theory of the LPM effect and experimental data. We believe that this allows
one to understand the origin of the normalization problem for
high-$Z$ targets discussed in \cite{SLAC}.

\begin{table}
\caption{\label{table1} The dependence of the relative Coulomb correction
$-\delta_{\scriptscriptstyle CC}\!\left[\langle
dI/d\omega\rangle\right]$  ($\%$) on the Migdal parameter $s$ in the
regime of small LPM suppression for some
high-$Z$ targets.}
\bigskip

\begin{center}
\begin{tabular}{ccccccccc}
\br
Target&$Z$&s=1.0 &s=1.1 &s=1.2 &s=1.3&s=1.5&s=2.0&s=\,$\infty$\\
\mr
Au&79&4.32&4.28&4.26&4.24&4.22&4.21&4.19\\
Pb&82&4.58&4.54&4.51&4.49&4.47&4.46&4.45\\
U&92&5.45&5.41&5.36&5.34&5.33&5.31&5.30\\
\br

\end{tabular}
\end{center}
$^a\,\bar\delta_{\scriptscriptstyle CC}\!\left[\left\langle
dI/d\omega\right\rangle\right]=(-4.50\pm 0.05)\%$ ($Z=82$),
$\bar\delta_{\scriptscriptstyle CC}\!\left[\left\langle
dI/d\omega\right\rangle\right] =(-5.35\pm 0.06)\%$ ($Z=92$),
${\delta}_{\scriptscriptstyle
CC}\!\left[\left\langle dI/d\omega\right\rangle\right]_\mu = (-4.70\pm
0.49)\%$.
\end{table}

\subsection{Applying the improved Moli\`{e}re theory to the description of
an analogue of the LPM effect theory for a thin target}

\medskip

Application of the Moli\`{e}re multiple scattering theory to the
analysis of experimental data \cite{SLAC} for a thin target is based on the use of the following expression for the spectral radiation rate \cite{ShF}:
\begin{eqnarray}\label{averaged}
\,\left\langle\frac{dI}{d\omega}\right\rangle &= &\int
w_{\scriptscriptstyle
M}(\vartheta)\frac{dI(\vartheta)}{d\omega}d^2\vartheta\, ,\nonumber\\
\frac{dI(\vartheta)}{d\omega}&=&\frac{2e^2}{\pi}
\left[\frac{2\chi^2+1}{\chi\sqrt{\chi^2+1}}
\ln\left(\chi+\sqrt{\chi^2+1}\right)-1\right]\ ,
\end{eqnarray}
which has two simple asymptotes at the small
and large values of the parameter $\chi=\gamma\vartheta/2$,
\begin{eqnarray}\label{asymptaver}
\left\langle\frac{dI}{d\omega}\right\rangle
&=&\frac{2e^2}{3\pi}\left\{\begin{array}{cl}
\gamma^2\overline{\vartheta^2},&\gamma^2\overline{\vartheta^2}\ll 1\;,\\
3\left[\ln
(\gamma^2\overline{\vartheta^2})-1\right],&\gamma^2\overline{\vartheta^2}\gg
1\;.\end{array}\right.
\end{eqnarray}
In the first case $\gamma^2\overline{\vartheta^2}\ll 1$, we have
\begin{eqnarray}\label{limcorrect1}
\delta_{\scriptscriptstyle CC}\!\left[\left\langle
\frac{dI}{d\omega}\right\rangle\right]&=&\delta_{\scriptscriptstyle
CC}\!\left[\left( \frac{dI}{d\omega}\right)_0\right]
\:=\:\frac{f(\xi)}{1-B^{\scriptscriptstyle B}}\ .
\end{eqnarray}
In the second case $\gamma^2\overline{\vartheta^2}\gg 1$, we find
\begin{eqnarray}
\label{16}
\Delta_{\scriptscriptstyle CC}\!\left[\ln\left(
\gamma^2\overline{\vartheta^2}\right)-1\right]&=&\Delta_{\scriptscriptstyle
CC}\!\left[\ln\left(\overline{\vartheta^2}\right)\right]
=\Delta_{\scriptscriptstyle CC}\big[\ln\left(B\right)\big]\ ,\\
\Delta_{\scriptscriptstyle
CC}[\ln\left(B\right)]&=&\Delta_{\scriptscriptstyle CC}[B]+f(Z\alpha)=
\delta_{\scriptscriptstyle CC}[B]\ .\nonumber
\end{eqnarray}
The Coulomb correction (\ref{16}) then becomes
\begin{eqnarray}
\Delta_{\scriptscriptstyle CC}\!\left[\ln\left(
\gamma^2\overline{\vartheta^2}\right)-1\right]&=&\frac{\delta_{\scriptscriptstyle
CC}[B]}{\left[\ln
(\gamma^2\overline{\vartheta^2})^{\scriptscriptstyle B}-1\right]}\ ,
\end{eqnarray}
and we arrive at a result:
\begin{eqnarray}\label{limcorrect2}
\delta_{\scriptscriptstyle CC}\!\left[\left\langle
\frac{dI}{d\omega}\right\rangle\right]&=& \frac{\Delta_{\scriptscriptstyle
CC}[\ln\big(\theta_a^{\,\prime}\big)]}{\left[\ln
(\gamma^2\overline{\vartheta^2})^{\scriptscriptstyle
B}-1\right]\bigg(1-B^{\scriptscriptstyle B}\bigg)}\ .
\end{eqnarray}

The numerical values of our corrections additionally improve the agreement between the predictions of the LPM effect theory analogue for a thin layer of matter and experiment \cite{SLAC} (figure 2, the dashed line "VKT") \cite{VKT}.

\begin{figure}[h!]

\begin{center}

\includegraphics[width=0.55\linewidth]{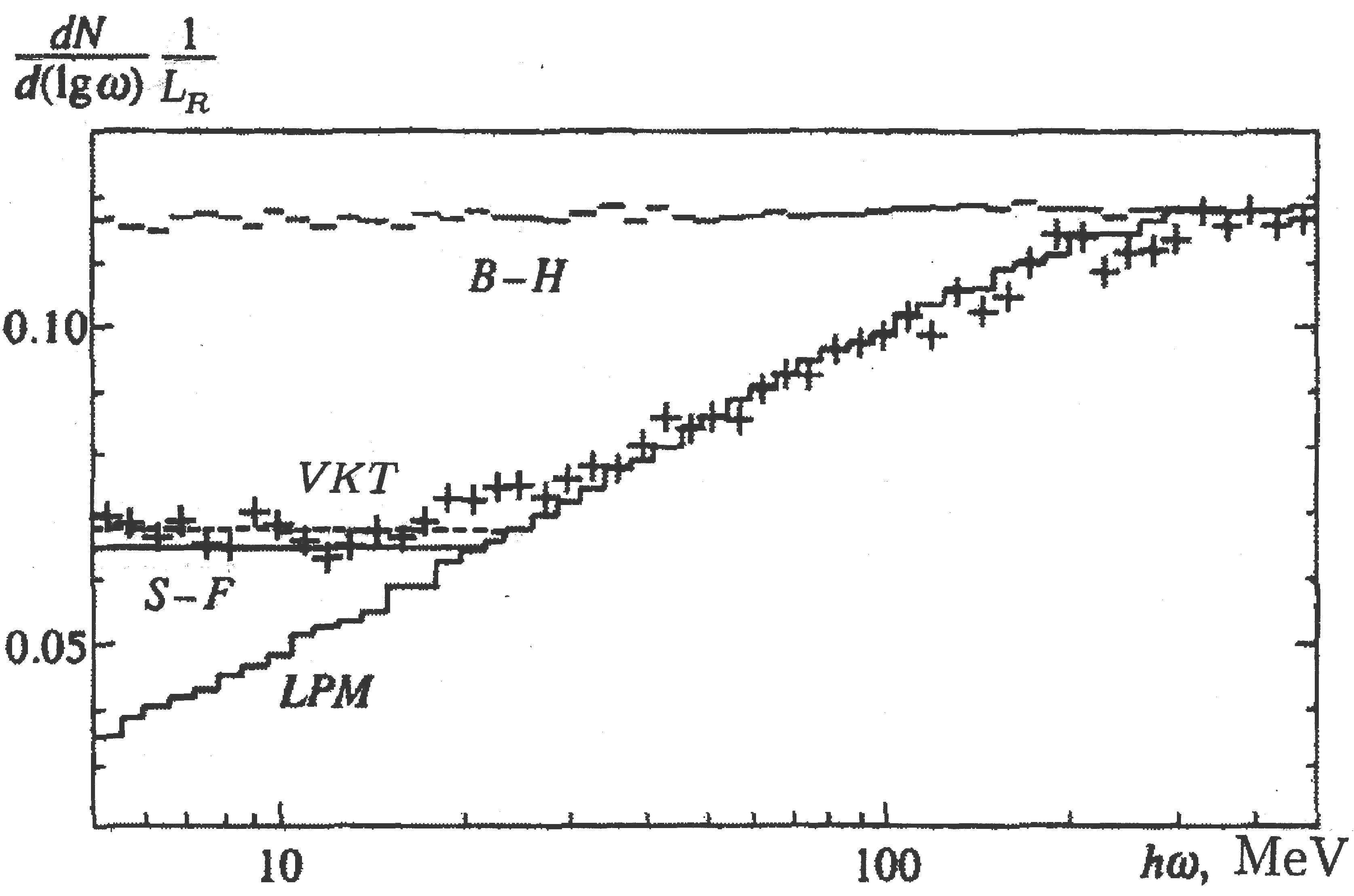}
\caption{ Measurement of the LPM effect over the range $30 < \omega < 300$ MeV and its analogue in the range $5 < \omega < 30$ MeV for the $0.7\% L_R$ gold target and 25 GeV electron beam. The signs `+' denote the experimental data; the histograms B--H and LPM give the Bethe--Heitler and the LPM Monte Carlo predictions \cite{SLAC}. The solid and dashed lines over the range  $\omega < 30$ MeV are the results of calculations without (S--F) and with the obtained Coulomb corrections (VKT) \cite{VKT}.}
\end{center}
\end{figure}

\subsection{Coulomb corrections to some quantities of quantum  LPM
effect theory}

\medskip
We have also obtained analytical and numerical results for  the Coulomb corrections
to the quantities of the quantum  Migdal LPM effect theory \cite{MQ,ASh}\footnote{Equations (\ref{migdal1}) and (\ref{Phi}) generalize (\ref{7}) and (\ref{8}) of the classical LPM effect theory for bremsstrahlung \cite{MC} to the case when the radiation recoil effect becomes significant.}

\begin{equation}\label{migdal1}
\left\langle\frac{dI}{d\omega}\right\rangle
=\frac{1}{4}\left(\frac{dI}{d\omega}\right)_0\Bigl\{\varepsilon^2 G(s)+2[1+(1-\varepsilon)^2]\Phi(s)\Bigl\}
\end{equation}
with the energy of radiation $\varepsilon=\omega/E$ in units of the incident particle energy $E$, Migdal's parameter $s=\sqrt{(\omega/q)}/8\gamma^2$ and the Migdal
functions $G(s)$ and $\Phi(s)$

\begin{eqnarray}\label{Phi}
G(s)&=&12\pi s^2-48s^2\sum_{k=0}^{\infty}\frac{1}{(k+s+1/2)^2+s^2}\, ,\nonumber
\end{eqnarray}
\begin{eqnarray}
\Phi(s)&=&6s-6\pi s^2+24s^2\sum_{k=1}^{\infty}\frac{2}{(k+s)^2+s^2}\,.
\end{eqnarray}
For the relative Coulomb correction to the Born spectral brems\-strah\-lung rate (\ref{migdal1}) one can get
\begin{eqnarray}
\delta_{\scriptscriptstyle
CC}\!\left[\left\langle \frac{dI}{d\omega}\right\rangle \right]=\delta_{\scriptscriptstyle
CC}\!\left[\left(\frac{dI}{d\omega}\right)_0\right]+\frac{\Delta_{\scriptscriptstyle
CC}[G(s)]+\Delta_{\scriptscriptstyle
CC}[\Phi(s)]}{\vert G^{\scriptscriptstyle B}(s)+\Phi^{\scriptscriptstyle B}(s)\vert}\nonumber
\end{eqnarray}
\begin{eqnarray}
~~~~~=\delta_{\scriptscriptstyle
CC}\!\left[\left(\frac{dI}{d\omega}\right)_0\right]
+\frac{G^{\scriptscriptstyle B}}{\vert G^{\scriptscriptstyle B}+\Phi^{\scriptscriptstyle B}\vert}\delta_{\scriptscriptstyle
CC}[G]
+\frac{\Phi^{\scriptscriptstyle B}}{\vert G^{\scriptscriptstyle B}+\Phi^{\scriptscriptstyle B}\vert}\delta_{\scriptscriptstyle
CC}[\Phi]\, ,\nonumber
\end{eqnarray}
\begin{eqnarray}
\label{21}
\delta_{\scriptscriptstyle
CC}\!\left[\left(\frac{dI}{d\omega}\right)_0\right]=
\frac{f(\xi)}{1-B^{\scriptscriptstyle B}}\ .
\end{eqnarray}

\newpage

\noindent For the regime of small LPM suppression \cite{MQ,ASh}
\begin{eqnarray}
G(s)_{s\to \infty}&\to& 1-0.029/s^4\, ,\nonumber\\
\Phi(s)_{s\to \infty}&\to& 1-0.012/s^4\, ,
\end{eqnarray}
in accordance with (\ref{rel})
and (\ref{12}), we have obtained:
\begin{eqnarray}
\label{23}
\delta_{\scriptscriptstyle \mathrm{CC}}\left[G(s) \right]
&=&0.029\,\frac{\delta_{\scriptscriptstyle \mathrm{CC}}\left[s^4
\right]}{\delta_{\scriptscriptstyle \mathrm{CC}}\left[s^4 \right]+1}
\cdot\frac{1}{\left(s^4\right)^{\scriptscriptstyle
\mathrm{B}}-0.029}\ ,\nonumber\\
\delta_{\scriptscriptstyle \mathrm{CC}}\left[\Phi(s) \right]
&=&0.012\,\frac{\delta_{\scriptscriptstyle \mathrm{CC}}\left[s^4
\right]}{\delta_{\scriptscriptstyle \mathrm{CC}}\left[s^4 \right]+1}
\cdot\frac{1}{\left(s^4\right)^{\scriptscriptstyle
\mathrm{B}}-0.012}\ ,\nonumber\\
&&\delta_{\scriptscriptstyle \mathrm{CC}}\left[s^4
\right]=1-\frac{1}{\delta^2_{\scriptscriptstyle
\mathrm{CC}}\big[\overline{\vartheta^2}\big]+2\delta_{\scriptscriptstyle
\mathrm{CC}}\big[\overline{\vartheta^2}\big] +1}\,,\nonumber\\
&&\delta_{\scriptscriptstyle
\mathrm{CC}}\big[\overline{\vartheta^2}\big] =
\frac{\Delta_{\scriptscriptstyle
CC}[\ln\big(\theta_a^{\,\prime}\big)]}{1-B^{\scriptscriptstyle B}}=
\frac{f(\xi)}{1-B^{\scriptscriptstyle B}}\, .
\end{eqnarray}

\medskip

\begin{table}
\noindent {\bf Table 2.} The dependence of the relative Coulomb correction
$-\delta_{\scriptscriptstyle CC}\!\left[\langle
dI/d\omega\rangle\right]$ ($\%$) on the Migdal parameter $s$ in the
regime of small LPM suppression for some
high-$Z$ targets.
\begin{center}
{\bf 1.} $B^{B}=8.46$, $s=1.5$ and $\beta =1$.

\medskip
\begin{tabular}{ccccccc}
\br
Target&$Z$~~~~&$\delta_{\scriptscriptstyle CC}\!\left[s^4\right] $&$\delta_{\scriptscriptstyle CC}\!\left[G(s)\right] $&$\delta_{\scriptscriptstyle CC}\left[\Phi(s)\right]$&$\delta_{\scriptscriptstyle
CC}\left[\left(\frac{dI}{d\omega}\right)_{\scriptscriptstyle
0}\right]$ &$\delta_{\scriptscriptstyle
CC}\left[\left\langle\frac{dI}{d\omega}\right\rangle\right]$\\[.2cm]
\mr
 W~~&74~~~~&$-0.0799$&$-0.0004$ &~$-0.0002$& $-0.0377$   &$-0.0380$\\
Au~~&79~~~~&$-0.0896$&$-0.0005$&~$-0.0002$&$-0.0419$   &$-0.0422$\\
Pb~~&82~~~~&$-0.0953$&$-0.0005$&~$-0.0002$&$-0.0445$   &$-0.0448$\\
 U~~&92~~~~&$-0.1148$&$-0.0006$ &~$-0.0002$& $-0.0529$   &$-0.0533$\\
 \br
\end{tabular}
\medskip

{\bf 2.} $B^{B}=4.50$, $s=1.5$ and $\beta =1$.

\medskip
\begin{tabular}{ccccccc}
\br
Target&$Z$~~~~&$\delta_{\scriptscriptstyle CC}\!\left[s^4\right] $&$\delta_{\scriptscriptstyle CC}\!\left[G(s)\right] $&$\delta_{\scriptscriptstyle CC}\left[\Phi(s)\right]$&$\delta_{\scriptscriptstyle
CC}\left[\left(\frac{dI}{d\omega}\right)_{\scriptscriptstyle
0}\right]$ &$\delta_{\scriptscriptstyle
CC}\left[\left\langle\frac{dI}{d\omega}\right\rangle\right]$\\[.2cm]
\mr
 W~~&74~~~~&$-0.1822$&$-0.0009$ &~$-0.0004$& $-0.0803$   &$-0.0810$\\
Au~~&79~~~~&$-0.2060$&$-0.0009$&~$-0.0004$&$-0.0894$   &$-0.0901$\\
Pb~~&82~~~~&$-0.2207$&$-0.0010$&~$-0.0004$&$-0.0949$   &$-0.0956$\\
 U~~&92~~~~&$-0.2707$&$-0.0012$ &~$-0.0005$& $-0.1129$   &$-0.1138$\\
 \br
\end{tabular}

\end{center}
\end{table}

Table 2 listed the results of
numerical estimation of the relative Coulomb corrections $\delta_{\scriptscriptstyle CC}\!\left[s^4\right] $,
$\delta_{\scriptscriptstyle CC}\left[G(s)
\right]$, $\delta_{\scriptscriptstyle CC}\left[\Phi(s)
\right]$ (\ref{23}), $\delta_{\scriptscriptstyle CC}\left[(dI/d\omega)_0\right]$ (\ref{10}) and $\delta_{\scriptscriptstyle CC}\left[\langle dI/d\omega\rangle\right]$ (\ref{21})
in the regime of small LPM suppression for some  high-$Z$
targets and  $B^{B}$ values from 8.46 \cite{SLAC} to 4.5 at $s=1.5$.
It shows
that while the modules of corrections $\delta_{\scriptscriptstyle CC}\!\left[s^4\right]$
for high-$Z$ absorbers reach the values about $11.5\%$ for $B^B = 8.46$ and $27\%$
for $B^B = 4.5$, the modules of $\delta_{\scriptscriptstyle CC}\!\left[G(s)\right] $ and
$\delta_{\scriptscriptstyle CC}\left[\Phi(s)\right]$ do not exceed
$0.1\%$. The dominant contribution to the correction $\delta_{\scriptscriptstyle
CC}\left[\left\langle dI/d\omega\right\rangle\right]$ gives
$\delta_{\scriptscriptstyle CC}\left[\left(dI/d\omega\right)_{\scriptscriptstyle
0}\right]$. It is seen from table~2 that $\delta_{\scriptscriptstyle
CC}\left[\left\langle dI/d\omega\right\rangle\right]\sim 5.33\%$ while $\delta_{\scriptscriptstyle
CC}\left[\left(dI/d\omega\right)_{\scriptscriptstyle
0}\right]\sim 5.29\%$ for $Z=92$ and $B^B = 4.5$;
$\delta_{\scriptscriptstyle
CC}\left[\left\langle dI/d\omega\right\rangle\right]\sim 11.4\%$ whereas $\delta_{\scriptscriptstyle
CC}\left[\left(dI/d\omega\right)_{\scriptscriptstyle
0}\right]\sim 11.3\%$ for $Z=92$ and $B^B = 4.5$.
Thereby, we have
$\delta_{\scriptscriptstyle
CC}\left[\left\langle dI/d\omega\right\rangle\right]\approx \delta_{\scriptscriptstyle
CC}\left[\left(dI/d\omega\right)_{\scriptscriptstyle
0}\right]\gg \delta_{\scriptscriptstyle CC}\!\left[G(s)\right]\approx 2\delta_{\scriptscriptstyle CC}\left[\Phi(s)\right]$ in the regime of small LPM suppression.


\newpage

\noindent For the regime of strong LPM suppression

\begin{equation}
G(s)_{s\to 0}\rightarrow 12\pi s^2,\; \Phi(s)_{s\to 0}\rightarrow 6s,
\end{equation}
we find
\begin{eqnarray}
\label{25}
\delta_{\scriptscriptstyle \mathrm{CC}}\left[G(s) \right]
&=&1-\frac{1}{\delta_{\scriptscriptstyle
CC}\big[\overline{\vartheta^2}\big]+1}\ ,\nonumber\\
\delta_{\scriptscriptstyle \mathrm{CC}}\left[\Phi(s) \right]
&=&1- \frac{1}{\sqrt{\delta_{\scriptscriptstyle
CC}\big[\overline{\vartheta^2}\big]+1}}\ .
\end{eqnarray}\\

Table~3 and figure~3 demonstrate the $Z$ dependence of the corrections $\delta_{\scriptscriptstyle CC}\!\left[G(s)\right]$,  $\delta_{\scriptscriptstyle CC}\!\left[\Phi(s)
\right]$ (\ref{25}), $\delta_{\scriptscriptstyle CC}\!\left[(dI/d\omega)_0\right]$ (\ref{10}) and
$\delta_{\scriptscriptstyle CC}\!\left[\langle dI/d\omega\rangle\right]$ (\ref{21})
in the regime of strong LPM suppression over the ranges $0\leq Z \leq 100$ and $4.5\leq B^B \leq 8.5$ at $s=0.05$. It can be seen from table 3  that $\delta_{\scriptscriptstyle CC}\!\left[G(s)
\right]$ is comparable with $\delta_{\scriptscriptstyle CC}\!\left[(dI/d\omega)_0\right]$ in the case of strong suppression. The value of
$\big\vert\delta_{\scriptscriptstyle CC}\left[G(s)
\right]\big\vert$  is twice as large than
$\big\vert\delta_{\scriptscriptstyle CC}\left[\Phi(s)
\right]\big\vert$ and reaches about $4.7\%$ at $B^B=8.46$ and $10.5\%$ at $B^B=4.50$
for $Z=82$ (Pb). In other words,
$\delta_{\scriptscriptstyle
CC}\!\left[\left(dI/d\omega\right)_{\scriptscriptstyle
0}\right]\approx\delta_{\scriptscriptstyle CC}\!\left[G(s)\right]\approx
2\delta_{\scriptscriptstyle CC}\!\left[\Phi(s)\right] $, and $\delta_{\scriptscriptstyle
CC}\!\left[\left\langle dI/d\omega\right\rangle\right]\approx 1.7\delta_{\scriptscriptstyle
CC}\!\left[\left(dI/d\omega\right)_{\scriptscriptstyle 0}\right]$ 
at $s\ll 1$.
The  $\delta_{\scriptscriptstyle
CC}\!\left[\left\langle dI/d\omega\right\rangle\right]$ module amounts to $7.4\%$
at $B^B=8.46$ and $16.0\%$ at $B^B=8.46$ for $Z=82$ ($s=0.05$).
Its upper limit is about $19\%$
over the ranges considered.

\begin{table}
\noindent {\bf Table 3.} Coulomb corrections  to the quantities
$\delta_{\scriptscriptstyle CC}\!\left[G(s)
\right]$, $\delta_{\scriptscriptstyle CC}\!\left[\Phi(s)
\right]$, $\delta_{\scriptscriptstyle CC}\!\left[(dI/d\omega)_0\right]$
and $\delta_{\scriptscriptstyle CC}\!\left[\langle dI/d\omega\rangle\right]$
 of the quantum Migdal LPM in the regime of strong LPM suppression.
\begin{center}
{\bf 1.} $B^{B}=8.46$, $s=0.05$ and $\beta =1$.

\medskip
\begin{tabular}{ccccccc}
\br
Target&~$Z$~~~~&$f(\xi)$&$\delta_{\scriptscriptstyle CC}\!\left[G(s)\right] $&$\delta_{\scriptscriptstyle CC}\left[\Phi(s)\right]$&$\delta_{\scriptscriptstyle
CC}\left[\left(\frac{dI}{d\omega}\right)_{\scriptscriptstyle
0}\right]$ &$\delta_{\scriptscriptstyle
CC}\left[\left\langle\frac{dI}{d\omega}\right\rangle\right]$\\[.2cm]
\mr
 C~~&6~~~ &0.0041&$-0.0005$ &~$-0.0003$& $-0.0005$   &$-0.0008$\\
Al~~&13~~~~&0.0107&$-0.0014$&~$-0.0007$&$-0.0014$   &$-0.0023$\\
Fe~~&26~~~~&0.0430&$-0.0058$&~$-0.0029$&$-0.0058$   &$-0.0094$\\
 W~~&74~~~~&0.2810&$-0.0392$ &~$-0.0194$& $-0.0377$   &$-0.0621$\\
Au~~&79~~~~&0.3130&$-0.0438$&~$-0.0216$&$-0.0419$   &$-0.0698$\\
Pb~~&82~~~~&0.3320&$-0.0466$&~$-0.0230$&$-0.0445$   &$-0.0738$\\
 U~~&92~~~~&0.3950&$-0.0560$ &~$-0.0276$& $-0.0530$   &$-0.0887$\\
 \br
\end{tabular}
\medskip

{\bf 2.}  $B^{B}=4.50$, $s=0.05$ and $\beta =1$.

\medskip
\begin{tabular}{ccccccc}
\br
Target&~$Z$~~~~&$f(\xi)$&$\delta_{\scriptscriptstyle CC}\!\left[G(s)\right] $&$\delta_{\scriptscriptstyle CC}\left[\Phi(s)\right]$&$\delta_{\scriptscriptstyle
CC}\left[\left(\frac{dI}{d\omega}\right)_{\scriptscriptstyle
0}\right]$ &$\delta_{\scriptscriptstyle
CC}\left[\left\langle\frac{dI}{d\omega}\right\rangle\right]$\\[.2cm]
\mr
 C~~&6~~~ &0.0041&$-0.0012$ &~$-0.0006$& $-0.0012$   &$-0.0019$\\
Al~~&13~~~~&0.0107&$-0.0031$&~$-0.0015$&$-0.0031$   &$-0.0050$\\
Fe~~&26~~~~&0.0430&$-0.0125$&~$-0.0062$&$-0.0123$   &$-0.0201$\\
 W~~&74~~~~&0.2810&$-0.0873$ &~$-0.0427$& $-0.0803$   &$-0.1349$\\
Au~~&79~~~~&0.3130&$-0.0982$&~$-0.0479$&$-0.0894$   &$-0.1507$\\
Pb~~&82~~~~&0.3320&$-0.1049$&~$-0.0511$&$-0.0949$   &$-0.1603$\\
 U~~&92~~~~&0.3950&$-0.1273$ &~$-0.0617$& $-0.1129$   &$-0.1921$\\
 \br
\end{tabular}

\end{center}
\end{table}

\begin{figure}[h!]

\begin{center}

\includegraphics[width=0.35\linewidth]{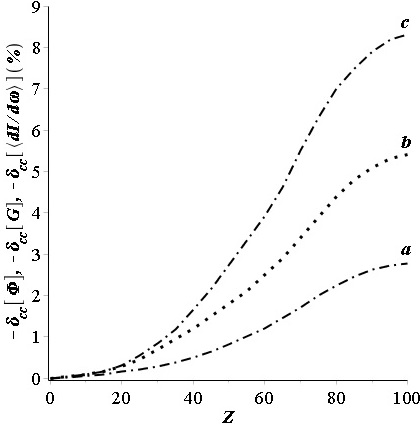}
\caption{The $Z$ dependence of relative CCs
to the quantities of the quantum LPM effect theory  for the regime of strong
LPM suppression: $(a)$  $\delta_{\scriptscriptstyle CC}\!\left[\Phi(s)\right]$,  $(b)$  $\delta_{\scriptscriptstyle CC}\!\left[G(s) \right]$ and $(c)$  $\delta_{\scriptscriptstyle
CC}\!\left[\left\langle dI/d\omega\right\rangle\right]$ at $s=0.05$ and $B^B=8.46$.
}
\end{center}
\end{figure}

\newpage
Thus, we can conclude that such corrections as  $\delta_{\scriptscriptstyle
CC}\left[\left\langle dI/d\omega\right\rangle\right]$, $\delta_{\scriptscriptstyle CC}\left[G(s)
\right]$ and $\delta_{\scriptscriptstyle CC}\left[(dI/d\omega)_0\right]$  become
significant in the regime of strong LPM suppression
and must be borne in mind, e.g., in the investigations of the LPM showers
in extremely high-energy region.

\section{Summary and outlook}
\medskip

\begin{itemize}
\item Within the quantum Migdal theory of the LPM effect, we have obtained the analytical
results for the Coulomb corrections to the Born bremsstrahlung rate $\left\langle dI/d\omega\right\rangle$ together with CCs to  the Migdal functions  $G(s)$ and $\Phi(s)$
in the regimes of small and strong LPM suppression based on results of the refined 
Moli\`{e}re multiple scattering theory. 

\item We evaluated the above correction in the regime of small LPM suppression over the ranges
$6\leq Z \leq 92$, $4.5\leq B^B \leq 8.46$, and $1.5\leq s \leq \infty$
and showed that while the modules of $\delta_{\scriptscriptstyle CC}\!\left[G(s)\right] $ and
$\delta_{\scriptscriptstyle CC}\left[\Phi(s)\right]$ do not exceed $0.1\%$, the modulus of relative Coulomb
correction to the Born bremsstrahlung rate $\delta_{\scriptscriptstyle
CC}\!\left[\left\langle dI/d\omega\right\rangle\right]$ can reach a value of about $11\%$ at 
$B^B=4.5$ and $Z=92$.

\item We have performed analogous calculations for the regime of strong LPM suppression
over the ranges $6\leq Z \leq 92$ and $4.5\leq B^B \leq 8.46$ at $s=0.05$
and found that  $\delta_{\scriptscriptstyle
CC}\!\left[\left\langle dI/d\omega\right\rangle\right]$ has a sufficiently large value that
ranges from around $-8\%$ for $B^B=8.46$  up to $-19\%$ for $B^B=4.5$ ($Z=92$).
The contribution of this correction should be appropriately considered, for instance, in the accurate modeling  of electromagnetic cascade  LPM showers in ultrahigh-energy region.

\item The further development of this approach involves primarily taking into account
    together with the bremsstrahlung also pair production.
On the basis of this analytical approach and the Monte-Carlo methods including
refined quantities of the quantum LPM effect theory,
we will try to contribute
to the clarification for the structure of three dimensional cascade showers
and also to solving the problem of diversity among them \cite{MA}.

\item  The developed approach can also be useful for the analysis of
cosmic-ray experiments in ultrahigh-energy region, where the LPM effect becomes significant (e.g., in applications motivated by a research of the superhigh-energy IceCubes neutrino-induced showers over the energy region $10^{16}\leq E \leq 10^{19}$ \cite{IceCube},
in exploring the properties of the extremely high-energy LPM showers, for compu\-ting
their characte\-ristics,
etc. \cite{KATM}).

\end{itemize}

\section*{Acknowledgments}
\medskip

The author would like to thank
 Prof.~Akeo Misaki (Saitama University, Japan) for his
 interest to the work and stimulating discussions.

\end{document}